\documentclass[twocolumn,twoside,slac_two]{revtex4}
\usepackage{graphicx}
\usepackage{fancyhdr}
\usepackage{hyperref}
\usepackage{natbib}

\def\sqig{$\sim$}

\def\degrees{$^{\circ}$}
\newcommand\aap{{A\&A}}%
\newcommand\apj{{ApJ}}%
\newcommand\apjl{{ApJ}}%
\newcommand\apjs{{ApJS}}%
\newcommand\planss{{Planet.~Space~Sci.}}%

\begin{document}

\title{Looking for Stars and Finding the Moon:\\
Effects of Lunar Gamma-ray Emission on \textit{Fermi} LAT Light Curves}

%


\author{Robin Corbet}
\affiliation{University of Maryland, Baltimore County, and
Code 662 NASA Goddard Space Flight Center,
Greenbelt Rd., MD 20771,
USA}
\author{C.C. Cheung}
\affiliation{Space Science Division, Naval Research Laboratory,
Washington, DC 20375-5352,
USA}

\author{Matthew Kerr}
\affiliation{Department of Physics and SLAC National Accelerator Laboratory,
Stanford University, Stanford, CA 94305,
USA}

\author{Paul S. Ray}
\affiliation{Space Science Division, Naval Research Laboratory,
Washington, DC 20375-5352,
USA}

\begin{abstract}
We are conducting a search for new gamma-ray binaries by making high signal-to-noise light curves of all
cataloged \textit{Fermi} LAT sources and searching for periodic variability using appropriately weighted power spectra.  
The light curves are created using a variant of aperture photometry where photons are weighted by the probability 
that they came from the source of interest.  From this analysis we find that the light curves of a number of sources
near the ecliptic plane are contaminated by gamma-ray emission from the Moon. This shows itself as modulation on 
the Moon's sidereal period in the power spectra. We demonstrate that this contamination can be removed by excluding 
times when the Moon was too close to a source. We advocate that this data screening should generally be used 
when analyzing LAT data from a source located close to the path of the Moon.
\end{abstract}

\maketitle

\thispagestyle{fancy}


\section{Introduction: Hunting for Gamma-ray Binaries}
At X-ray energies, the extra-solar sky is dominated by the emission from accreting binary systems containing black holes 
and neutron stars. However, at higher energies (GeV to TeV) very few interacting neutron star or
black hole binaries are known sources~\citep{Hill2011}.
The relative paucity of gamma-ray binaries can be attributed to the necessity for not only a power supply, 
but also non-thermal mechanisms~\citep{Dubus2006, Mirabel2006}. 
There are, however, evolutionary reasons to expect more gamma-ray binaries to exist~\citep{Meurs1989}, 
and there are many unidentified \textit{Fermi} LAT sources.
Gamma-ray binaries are expected to show orbitally-modulated gamma-ray emission due to 
changes in viewing angle and, in eccentric orbits, the degree of the binary interaction. 
Periodic modulation has indeed been seen in LS 5039 (3.9 day period), LS I +61 303 (26.5 days), 
Cygnus X-3 (4.8 hours)~\citep{Hill2011}, and 1FGL J1018.6-5856 (16.65 days)~\citep{Corbet2011, Fermi2012} 
and emission is orbital phase dependent 
for PSR B1259-63 (3.4 years)~\citep{Abdo2011}. A search for periodic modulation of gamma-ray flux from \textit{Fermi} LAT sources
may thus yield further gamma-ray binaries, potentially revealing the predicted HMXB precursor population. 
The second \textit{Fermi} LAT catalog of gamma-ray sources (``2FGL''~\citep{Nolan2012}) contains 1873 sources, many  of which do not 
have confirmed counterparts at other wavelengths and thus are potentially gamma-ray binaries.

In order to search for modulation we regularly update 0.1 - 200 GeV light curves for all 2FGL sources
and calculate power spectra of these. We use aperture photometry with a 3\degrees\ radius, with photons weighted by the probability 
that they came from the source of interest to increase the signal-to-noise level. 
To avoid solar gamma-ray emission, we exclude times when the Sun was closer than 5\degrees\
to an aperture.
We then calculate power 
spectra of all light curves to search for periodic modulation. To account for variations in exposure, 
each time bin’s contribution to the power spectrum is weighted by its relative exposure~\citep{Fermi2012}.

\section{Complex Modulation Patterns in Two \textit{Fermi} Sources}
\begin{figure}
\includegraphics[angle=-90,width=83mm]{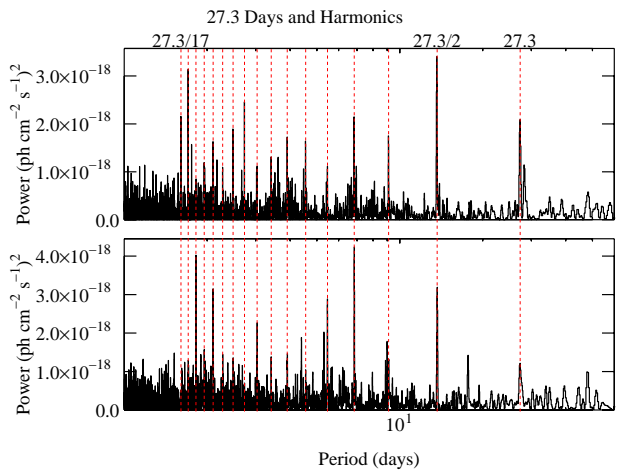}%
\caption{Both 2FGL J2356.3+0432 (top) and 2FGL J0753.2+1937 (bottom) 
show a complex pattern of peaks in their power spectra. These peaks are harmonics of a 27.3 day period - 
up to the 17$^{\rm th}$ harmonic is detectable.}\label{fig-power}
\end{figure}

\begin{figure}
\includegraphics[angle=-90,width=83mm]{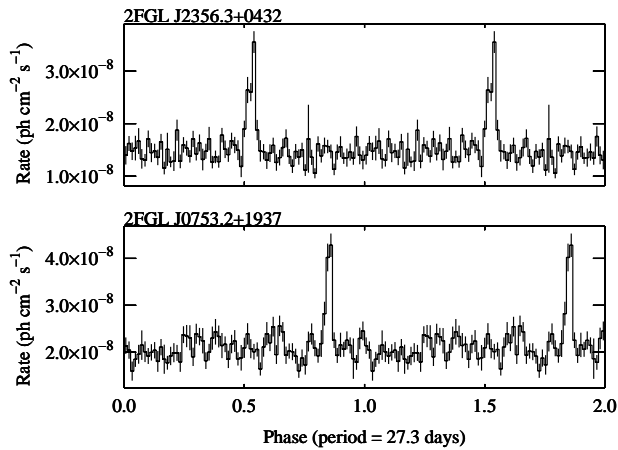}%
\caption{Modulation for both  2FGL J2356.3+0432 and 2FGL J0753.2+1937 is caused by sharp ``flares'' that recur with a 27.3 day period, 
but with different epochs of maximum flux.}\label{fig-fold}
\end{figure}

From an examination of the power spectra for all sources in the 2FGL catalog, orbital modulation is strongly detected
in the known gamma-ray binaries LS 5039, LS I +61 303, and 1FGL J1018.6-5856 (= 2FGL J1019.0-5856).
Artifacts near 1 day and the precession period of \textit{Fermi}'s orbit at \sqig53 days are also
seen in the power spectra of a number of 
sources\footnote{\url{http://fermi.gsfc.nasa.gov/ssc/data/analysis/LAT_caveats_temporal.html}}. 
In addition to these, we noted complex
sets of peaks in the power spectra of 2FGL J0753.2+1937 and 2FGL J2356.3+0432. 
2FGL J0753.2+1937 does not have an identified counterpart at other wavelengths, 
while 2FGL J2356.3+0432 is identified with the blazar MG1 J235704+0447~\citep{Nolan2012}.
Although these two sources
are widely separated on the sky, it was determined that the peaks in both sources
were all harmonics of a 27.3 day period (Figure~\ref{fig-power}).
When the light curves are folded on this period, brief flares are seen in both sources (Figure~\ref{fig-fold}).

\section{Lunar Gamma-Rays}
Interactions of cosmic rays with the Moon's surface result in the production of gamma rays. 
This makes the Moon a rather bright source for the \textit{Fermi} LAT with a flux above 100 MeV 
of \sqig10$^{-6}$ ph cm$^{-2}$ s$^{-1}$~\citep{Abdo2012}, 
and it was even detectable with EGRET~\citep{Thompson1997}. We note that the Sun is 
also a gamma-ray source. Although the 2FGL catalog notes sources potentially affected by solar emission, 
no such analysis was done for the Moon~\citep{Nolan2012}.

The Moon's sidereal period is 27.321 days. The sharp recurrent flares from 2FGL J0753.2+1937 and 2FGL J2356.3+0432
can be understood as due to repeated passages of the Moon sufficiently close to these sources to affect the light curves.

\section{Optimizing Lunar Detection: Summed Harmonics}
\begin{figure}
\includegraphics[angle=-90,width=83mm]{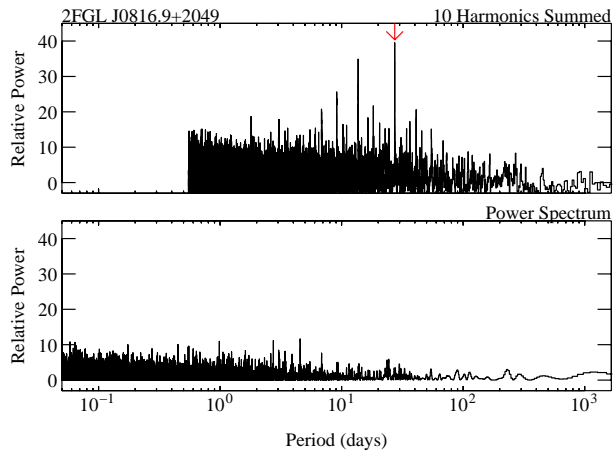}%
\caption{Lunar contamination of 2FGL J0816.9+2049 is not directly detected in the power spectrum (bottom). However, the summed harmonic 
modification of this (top) clearly shows a 27.3 day period due to the Moon.}\label{fig-harm}
\end{figure}

Power spectra are not ideal for detecting brief flaring activity as this strongly 
non-sinusoidal modulation results in the power being spread over a very large number of harmonics. 
We investigated other period-detection techniques such as Stellingwerf's phase dispersion minimization~\citep{Stellingwerf1978} 
method and others. It was found that lunar modulation was best detected by creating a modified power spectrum,
similar to the Z$^2$ test~\citep{Buccheri1983},
with each point replaced with a sum of itself and up to the 10th harmonic. This is illustrated in
Figure~\ref{fig-harm} where we show the power spectrum, and summed-harmonic power spectrum of 2FGL J0816.9+2049,
a source which is identified with the blazar 2FGL J0816.9+2049~\citep{Nolan2012}.
From harmonic-summed power spectra of the entire 2FGL catalog we detected 38 sources 
that suffered from significant lunar contamination (Table~\ref{table-sources}).

\begin{table}
\caption{\textit{Fermi} LAT Sources with Apparent Lunar Contamination}\label{table-sources}
\begin{tabular}{|l|r|r|}
\hline
\multicolumn{1}{|c}{\textbf{Source}} & \multicolumn{1}{|c}{\textbf{R.A. (J2000)}} & \multicolumn{1}{|c|}{\textbf{Decl. (J2000)}} \\
\multicolumn{1}{|c}{\textbf{}} & \multicolumn{1}{|c}{\textbf{(degrees)}} & \multicolumn{1}{|c|}{\textbf{(degrees)}} \\
\hline
2FGL J0009.0+0632  &  2.262  &  6.542\\
2FGL J0022.5+0607  &  5.643  &  6.124\\
2FGL J0023.5+0924  &  5.892  &  9.407\\
2FGL J0114.7+1326  & 18.675  &  13.441 \\
2FGL J0257.9+2025c & 44.480  &  20.423 \\
2FGL J0322.0+2336  & 50.516  &  23.611  \\
2FGL J0326.1+2226  & 51.536  &  22.439  \\
2FGL J0440.5+2554c & 70.146  &  25.903 \\
2FGL J0709.0+2236  & 107.274  &  22.600  \\
2FGL J0725.6+2159  & 111.400  &  21.990  \\
2FGL J0753.2+1937  & 118.320  &  19.623  \\
2FGL J0816.9+2049  & 124.250  &  20.823  \\
2FGL J0839.4+1802  & 129.863  &  18.036  \\
2FGL J0913.0+1553  & 138.251  &  15.893  \\
2FGL J0923.5+1508  & 140.895  &  15.138  \\
2FGL J0946.5+1015  & 146.648  &  10.259  \\
2FGL J1007.7+0621  & 151.932  &  6.353  \\
2FGL J1016.0+0513  & 154.014  &  5.229  \\
2FGL J1018.6+0531  & 154.659  &  5.524  \\
2FGL J1040.7+0614  & 160.182  &  6.246  \\
2FGL J1058.4+0133  & 164.615  &  1.566  \\
2FGL J1059.0+0222  & 164.767  &  2.374  \\
2FGL J1107.5+0223  & 166.878  &  2.386  \\
2FGL J1221.4$-$0633  & 185.358  &  $-$6.553   \\
2FGL J1256.5$-$1145  & 194.139  &  $-$11.753  \\
2FGL J1318.9$-$1228  & 199.745  &  $-$12.476  \\
2FGL J1424.2$-$1752  & 216.054  &  $-$17.880  \\
2FGL J1544.1$-$2554  & 236.042  &  $-$25.912\\
2FGL J1553.2$-$2424  & 238.322  &  $-$24.404  \\
2FGL J2000.8$-$1751  & 300.217  &  $-$17.857  \\
2FGL J2006.9$-$1734  & 301.734  &  $-$17.582  \\
2FGL J2031.4$-$1842  & 307.868  &  $-$18.703  \\
2FGL J2108.6$-$1603  & 317.159  &  $-$16.062  \\
2FGL J2120.6$-$1301  & 320.152  &  $-$13.030  \\
2FGL J2124.0$-$1513  & 321.023  &  $-$15.223  \\
2FGL J2154.0$-$1138  & 328.503  &  $-$11.634  \\
2FGL J2225.6$-$0454  & 336.424  &  $-$4.901  \\
2FGL J2356.3+0432  & 359.091  &  4.541  \\
\hline
\end{tabular}
\end{table}

\begin{figure*}
\includegraphics[width=167mm,trim=0cm 0cm 0cm 0cm,clip=true]{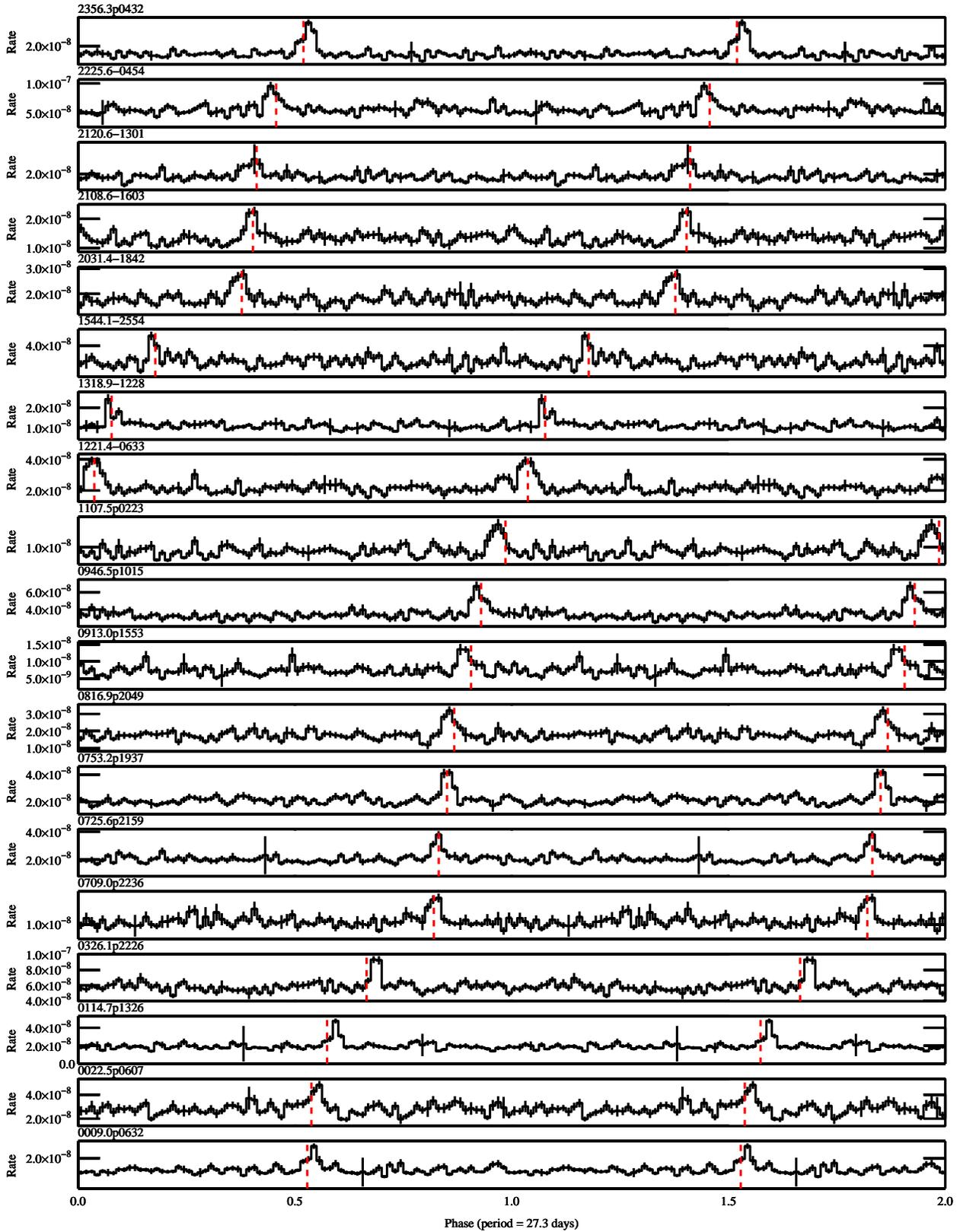}%
\caption{Light curves of selected \textit{Fermi} LAT sources from Table~\ref{table-sources} folded on the Moon's sidereal period. 
The vertical red lines are offset based only on the R.A. of each source and so roughly approximate the Moon's path.\label{fig-muti-fold}}
\end{figure*}

\section{Removing the Moon}
\textit{Fermi} spacecraft files do not currently include lunar coordinates. 
One of us (PSR) has provided a utility (``\texttt{moonpos}'') 
that uses the JPL SPICE toolkit~\citep{Acton1996}\footnote{\url{http://naif.jpl.nasa.gov/naif/toolkit.html}}
to add lunar coordinates. This is available from the \textit{Fermi} Science Support Center on the User Contributions 
web page\footnote{\url{http://fermi.gsfc.nasa.gov/ssc/data/analysis/user}}.

The addition of lunar coordinates enables filtering to exclude data obtained 
when the Moon was close to a source via the standard analysis tool \texttt{gtmktime}. We find that excluding data within 
8 degrees of a source removes almost all contamination.

\section{Applications to Searches for Flaring Binaries}
The technique of summing harmonics to reveal the presence of lunar contamination 
is also useful in the search for gamma-ray binaries. For example, the binary PSR B1259-63 
is only active for a brief portion of its 3.4 yr orbit~\citep{Abdo2011}. Other systems exhibiting similar repeating brief 
flares will be more readily detected using harmonically-summed power spectra. 
So far no definite non-lunar periodic flaring has been detected for any source, but the hunt continues.

\section{Summary}
\begin{itemize}
\setlength{\itemsep}{-12pt}
\setlength{\parsep}{0pt}
\item Lunar gamma-ray emission can significantly contaminate the light curves of LAT sources near the ecliptic plane.\\
\item Lunar modulation at 27.3 days is directly detected in the power spectra of a few sources.\\
\item Adding power-spectrum harmonics (\sqig10) reveals the 27.3 day signal in 38 2FGL sources.\\
\item Software has been developed to facilitate exclusion of lunar contaminated data. This is publicly available.
\end{itemize}

\noindent
We advocate:\\ 
(i) lunar proximity filtering should be done for any source close to the ecliptic.\\
(ii) lunar coordinates should be included in the standard \textit{Fermi} spacecraft files. 

The summed-harmonic technique is being used to search for gamma-ray binaries that briefly flare for only a short fraction of their orbit.

\begin{acknowledgments}
This work was partially supported by the NASA \textit{Fermi} Guest Observer Program (NNX12AH82G).
\end{acknowledgments}

\bigskip 

\end{document}